\documentclass[a4]{article}
\usepackage{graphicx}

\def\smallskip{\vskip3pt}
\def\medskip{\vskip6pt}
\def\bigskip{\vskip12pt}

\font\scap=cmcsc10

\font\tenmsb=msbm10
\font\sevenmsb=msbm7
\font\fivemsb=msbm5
\newfam\msbfam
\textfont\msbfam=\tenmsb
\scriptfont\msbfam=\sevenmsb
\scriptscriptfont\msbfam=\fivemsb
\def\Bbb#1{{\fam\msbfam\relax#1}}

\newcount\eqnumber
\def\neweq{{\rm{(\the\eqnumber)}}\global\advance\eqnumber by 1}
\def\eqdef#1{\eqno\xdef#1{\the\eqnumber}\neweq}
\def\newaeq{{\rm{(\the\eqnumber a)}}\global\advance\eqnumber by 1}
\def\eqdaf#1{\eqno\xdef#1{\the\eqnumber}\newaeq}
\def\eqdisp#1{\xdef#1{\the\eqnumber}\neweq}
\def\eqdasp#1{\xdef#1{\the\eqnumber}\newaeq}

\newcount\refnumber
\def\newref{{\the\refnumber}\global\advance\refnumber by 1}
\def\refdef#1{{\xdef#1{\the\refnumber}}\newref}

\begin{document}

\centerline{\bf Calculating algebraic entropies: an express method}
\bigskip
\medskip{\scap A. Ramani} and {\scap B. Grammaticos}
\quad{\sl IMNC, Universit\'e Paris VII \& XI, CNRS, UMR 8165, B\^at. 440, 91406 Orsay, France}
\medskip{\scap R. Willox} and {\scap T. Mase}\quad
{\sl Graduate School of Mathematical Sciences, the University of Tokyo, 3-8-1 Komaba, Meguro-ku, 153-8914 Tokyo, Japan }
\bigskip
{\sl Abstract}
\smallskip
We describe a method for investigating the integrable character of a given three-point mapping, provided that the mapping has confined singularities. Our method, dubbed ``express'', is inspired by a novel approach recently proposed by R.G. Halburd. While the latter aims at computing the exact degree growth of a given mapping based on the structure of its singularities, we content ourselves with obtaining an answer as to whether a given system is integrable or not. We present several examples illustrating our method as well as its limitations. We also compare the present method to the full-deautonomisation approach we recently introduced. 

\bigskip
1. {\scap Introduction}
\medskip
In his seminal paper [\refdef\veselov], ``Growth and integrability in the dynamics of mappings'', Veselov summarised the question of the detection of discrete integrability in the sentence: ``the integrability has an essential correlation with the weak growth of certain characteristics''. This aphorism acquires a real value once one makes more precise the `characteristics' in question. The one  put forward by Veselov is complexity, a notion first introduced by Arnold [\refdef\arnold]. Put simply, the complexity of a mapping of the plane  can be defined as the number of intersection points of a fixed curve with the image of a second curve under the iteration of the mapping. This notion of complexity lies at the origin of what Bellon and Viallet [\refdef\bellon] called algebraic entropy. Their approach, tailored to rational mappings, consists in introducing homogeneous coordinates and computing the common degree $d_n$ of the homogeneous polynomials that appear at the $n$th iteration of the mapping. They made their criterion more precise by considering the limit of the degree when $n\to\infty$, introducing the quantity
$$\varepsilon=\lim_{n\to\infty}{1\over n}\log d_n,$$
which they dubbed algebraic entropy. (The value of $\lambda=e^{\varepsilon}$ is often referred to as the dynamical degree of the mapping). A strictly positive value for $\varepsilon$ (corresponding to a dynamical degree greater than 1) is an indication of non-integrability, while integrable mappings have  zero algebraic entropy (and dynamical degree equal to 1). The latter statement means that in integrable mappings the iterations induce algebraic simplifications which lower substantially the degree of the iterates, up to the point that it becomes polynomial, while in the absence of such simplifications the degree would exhibit exponential growth. Notice that in the Bellon-Viallet approach the value of $\varepsilon$ is obtained in a heuristic way, i.e. by ``guessing'' the behaviour of $d_n$ based on the explicit computation of a sufficient number of iterates. However, as we shall explain, in the case of three-point mappings, it is possible to obtain the value of the algebraic entropy rigorously, in a very simple way.

The origin of the simplifications we alluded to above can be found in the singularities which appear spontaneously during the iteration of the  mapping [\refdef\scae]. The notion of singularity confinement was in fact introduced [\refdef\sincon] as a discrete integrability criterion, based on the observation that, for integrable mappings, singularities which appear spontaneously will disappear after a certain, usually small, number of iterations. It is exactly the simplifications which make the singularities disappear that lower the degree $d_n$. In some cases, however, the simplifications that lead to confined singularities do not suffice [\refdef\hiv] to curb the exponential growth of the degree. The difficulty that the existence of non-integrable mappings with confined singularities presented to the detection of integrability was settled with the introduction of the full-deautonomisation [\refdef\redemp] approach. The latter consists in performing the singularity analysis of an extension of the mapping assuming that its parameters are functions of the independent variable and, moreover, by enriching the mapping through the addition of terms which do not modify the singularity pattern. As shown in [\redemp] this approach makes it possible to obtain exactly the value of the algebraic entropy of mappings with confined singularities, be they integrable or not. In [\refdef\mase, \refdef\redeem] we presented the algebro-geometric justification of the full-deautonomisation method. 

Note that it is precisely an algebro-geometric approach that first allowed for the rigorous computation of the value of the algebraic entropy  [\refdef\favre, \refdef\take]. However, performing the regularisation of a confining mapping, through a succession of blow-ups, might not be the most efficient approach if the only information one seeks is the value of the algebraic entropy. This would be a strong argument in favour of the full-deautonomisation approach as a practical method to be used in the detection of discrete integrability, where it not for two minor problems that may arise in some cases. First, the necessary calculations can become prohibitively bulky and, second, the process of enriching the mapping through the addition of terms is something that crucially depends on one's intuition, which may turn out to be insufficient. These arguments show that there is still a need for a rapid method which, without offering all the information one can obtain from the full-deautonomisation analysis,  easily yields the exact value of the algebraic entropy. Establishing such a method is precisely the aim of the present paper.
\bigskip
2. {\scap A summary of Halburd's method}
\medskip
Our approach is based on recent findings by R.G. Halburd [\refdef\rod], upon which we construct what we call the express method for the computation of algebraic entropy. 
The essence of Halburd's method (and also of the express one) is to study the iterates of a (second order) rational mapping as rational functions $f_n(z)$ of a single variable $z\in\bar{\Bbb C}={\Bbb C}\cup\{\infty\}$, introduced through the initial conditions $x_0$ and $x_1$, and to calculate the degree (in $z$) of each iterate as the number of preimages that a generic value $w\in\bar{\Bbb C}$ of the corresponding rational function possesses, i.e., by counting the number of solutions (in $\bar{\Bbb C}$) of $f_n(z)=w$, for arbitrary $n$. This counting of preimages is performed based on information that can be obtained from the singularity confinement analysis of the mapping.

Let us illustrate Halburd's method on the mapping:
$$x_{n+1}x_{n-1}=a{x_n-b\over x_n-1},\eqdef\eqi$$
(where $a\ne0,1$ and $b\neq 0,1,a$), which has been studied in detail in [\mase]. This mapping possesses two confining singularity patterns
$$\{1,\infty,a,0,b\},$$
$$\{b,0,a,\infty,1\}.$$
Moreover the following, non-periodic, pattern
$$\{\cdots, x,\infty,x',0,x'',\infty,\cdots\},$$
also exists, where $x$ is arbitrary and $x', x'',\cdots$ are values depending on $x$. Although it is not periodic, we call such a pattern {\sl cyclic} because of the cyclic appearance of particular values, in this case $0$ and $\infty$.

In order to obtain the growth of the mapping we shall study its iterates, starting from initial conditions: 
$$x_0\qquad{\rm and}\qquad x_1=z,\eqdef\init$$
for arbitrary $z\in\bar{\Bbb C}$ and for a generic value of $x_0$ (in the sense that $x_0$ does not satisfy any special relation). We then compute the degree $d_n$  in $z$ of the iterates $x_n$ of (\eqi) as the number of preimages $d_n(w)$, counting multiplicities, of any given value $w$ of $x_n$, considered as a rational function of $z$ on $\bar{\Bbb C}$. Note that this choice of initial conditions is different from that in [\rod] where, generically, both initial conditions involve the variable $z$. In our case only $x_1$ introduces a $z$ dependence and $x_0$ is treated as a mere parameter in the rational functions we obtain. This is the choice of initial conditions we shall always assume throughout this paper.

For example, iterating the mapping (\eqi) we find
$$x_2={a\over x_0}{z-b\over z-1},\eqdef\eqii$$
which means that the degree $d_2=d_2(w)$ is equal to 1 for any value $w$ for $x_2$. Iterating once more we obtain $d_3=2$, since $x_0$ is taken to be generic (and, as such, is different for example from $a$). Iterating further we can compute the growth of the mapping, but at the price of calculations that can be repeated only a small number of times. 

However, a much more effective method does exist. Following [\rod], let us see how certain values among those appearing in the singularity pattern can arise when iterating the mapping. For instance, what are the contributions to the number of preimages of $x_n=1$? Since $x_0$ is generic, either a value of 1 appears four steps after a value $b$ that opens a pattern of the form $\{b,0,a,\infty,1\}$, or it appears spontaneously at the present iteration, thereby itself opening a pattern of the form $\{1,\infty,a,0,b\}$. (And the same can be said for $b$, mutatis mutandis). We denote the number of {\sl spontaneous occurrences} of the values 1 and $b$ at step $n$ by $U_n$ and $B_n$ respectively. We have thus
$$d_n\equiv d_n(1)=U_n+B_{n-4},\eqdef\eqiii$$
and similarly
$$d_n\equiv d_n(b)=B_n+U_{n-4}.\eqdef\eqiv$$
In order to compute $d_n(\infty)$, or equivalently $d_n(0)$, we remark that an $\infty$ appears if $x$ was equal  to 1 one step before or equal to $b$ three steps before. Moreover the contribution of the values $\infty$ or 0 appearing in the cyclic pattern  should also be taken into consideration. Since the generic initial value $x_0$ is neither $\infty$ nor 0, there exist exactly two instances of the length-4 basic pattern of cyclic type, involving 0 and $\infty$, that start from a finite value:
$$\{x_0,\infty,x',0,\cdots\},$$
$$\{x_0,0,x',\infty,\cdots\}.$$
From these two patterns, we see that there is exactly one contribution to $\infty$ at odd $n$ and no contribution at even $n$. The same is true for 0. Given the form of the mapping these are the only contributions to $\infty$ and 0.
We have thus 
$$d_n\equiv d_n(0)=B_{n-1}+U_{n-3}+{1-(-1)^n\over2},\eqdef\eqv$$
and
$$d_n\equiv d_n(\infty)=B_{n-3}+U_{n-1}+{1-(-1)^n\over2}.\eqdef\eqvi$$
Since the system we have set up is symmetrical in $U$ and $B$ and since the initial conditions are such that $U$ and $B$ play the same role, we deduce that $U_n=B_n$. From (\eqiii) and (\eqv) we thus find 
$$d_n=U_n+U_{n-4}=U_{n-3}+U_{n-1}+{1-(-1)^n\over2},\eqdef\eqvii$$
leading to 
$$U_n-U_{n-1}-U_{n-3}+U_{n-4}={1-(-1)^n\over2},\eqdef\eqviii$$
the solution of which is formally
$$U_n= n^2/12-(-1)^n/8+\alpha n+\beta+\gamma j^{n}+\delta j^{2n},\eqdef\eqix$$
where $j=\exp{2i\pi\over3}$.
Using relation (\eqvii) we can compute the degree $d_n$ and determine the values of the constants $\alpha,\beta,\gamma,\delta$ from the knowledge of the first few iterates. We find
$$\alpha=1/3,\ \beta=17/72,\ \gamma=j^2/9,\ \delta=j/9,\eqdef\eqx$$
which leads to the expression for the degree
$$d_n={1\over36}(6n^2+17-9(-1)^n-4j^n-4j^{2n}).\eqdef\eqxi$$
This is in perfect agreement with the degrees of the iterates calculated directly: 0, 1, 1, 2, 3, 5, 6, 9, 11, 14, 17, 21, 24, 29, 33, $\cdots$. The polynomial growth of the degree means that the algebraic entropy is 0, in accordance with the integrable character of the mapping. 

Note that the quadratic dependence of the degree on $n$ is due precisely to the fact that, in (\eqviii), we have taken into account the contribution of the term $(1-(-1)^n)/2$, due to the cyclic patterns. However, had we neglected this contribution we would have still concluded that the mapping is integrable since the polynomial growth (which is tantamount to zero algebraic entropy) is conditioned by the absence of a characteristic root greater than 1 for the homogeneous part of equation (\eqviii).
Generally speaking, a confining rational mapping that is not periodic, something which is tacitly assumed throughout this paper, only has finitely many singularity patterns (always of finite length) and only a finite number of cyclic patterns (each consisting in the repetition of a finite length  basic pattern, as in the present example). For such a mapping, any particular value can therefore appear only in finitely many cyclic patterns. Moreover, given our initial conditions (\init), since the number of instances of any particular cyclic pattern is equal to the (finite) number of arbitrary values contained in a single (basic) cycle, the contribution of the cyclic patterns to the number of preimages of any value, for any iterate of such a mapping, is necessarily a bounded function of $n$. 

Hereafter we shall write relations where such bounded contributions from cyclic patterns are omitted. To this end we introduce the symbol $\simeq$ for an approximate equality which holds only up to some bounded terms, which are neglected.

Note that in our analysis above we have not considered the role played by the value $x_n=a$. From the singularity pattern, it is clear that a value $a$ appears when $x$ is equal to either 1 or $b$ two steps before, meaning that there is a contribution $B_{n-2}+U_{n-2}$ to $d_n(a)$. However, a value $a$ can {\sl a priori} appear also outside a singularity pattern. Since we already know the exact value of $d_n$, we compute the difference $d_n-B_{n-2}-U_{n-2}$ which turns out to be equal to ${(2-j^n-j^{2n})/3}$. In hindsight this is not surprising since a value $a$ that occurs outside a confined singularity pattern can only arise in a cyclic pattern. If we start from a generic $x_0$, when $x_1$ takes the value $a$, we find that this value appears periodically every three steps. Similarly if $x_1$ takes the value $(x_0-b)/(x_0-1)$ then $x_2=a$, whereupon $a$ again appears periodically, every three steps. So there are exactly two such cyclic patterns in which $a$ appears once at indices  $3n+1$ and $3n+2$, respectively, but never at $3n$. Using the approximate equality we introduced in the previous paragraph, together with the fact that $U$ and $B$ play the same role, we therefore have 
$$d_n\equiv d_n(a)\simeq 2U_{n-2}.\eqdef\aqxii$$
Combining (\aqxii) with (\eqiii) or with (\eqv) and omitting the term $(1-(-1)^n)/2$, we find two different homogeneous equations, namely 
$$U_n-2U_{n-2}+U_{n-4}\simeq0,\eqdef\aqxiii$$
and 
$$U_{n-1}-2U_{n-2}+U_{n-3}\simeq0.\eqdef\aqxiii$$
The characteristic equation of the homogeneous parts of these equations, interpreted as strict equalities, are $(\lambda+1)^2(\lambda-1)^2=0$ and $(\lambda-1)^2=0$ respectively. Note that neither coincides with the characteristic equation of (\eqviii), $(\lambda^2+\lambda+1)(\lambda-1)^2=0$. However, what  all three characteristic equations have in common is that they do not have a root greater than 1. Therefore, since the contribution of the cyclic patterns that was neglected in these calculations is always bounded, none of these relations can give rise to an exponential growth of the degree. \bigskip
3. {\scap The express method}
\medskip
From our analysis of Halburd's method it is clear that the integrable character of a confining mapping is associated with the fact that the characteristic equation of the homogeneous system we obtain does not have a root greater than 1. Thus, when the only thing one cares about is the value of the algebraic entropy, one can simply neglect the contribution of cyclic patterns that may exist for such a mapping. This is what we call the express method. 

We shall illustrate our approach in the case of the discrete Painlev\'e I [\refdef\fokas]:
$$x_{n+1}+x_{n-1}={a_n\over x_n}+{1\over x_n^2},\eqdef\eqxii$$
where $a_n$ satisfies the relation $a_{n+1}-2a_n+a_{n-1}=0$.
The confined singularity pattern is
$$\{0,\infty^2,0\}.$$
 Given the form of the right-hand side, infinite values of $x$ outside a confined pattern can only come from a cyclic pattern. 
We denote by $Z_n$ the spontaneous occurrences of 0 at some step $n$. The total number of zeros occurring at $n$ is the sum of $Z_n$ and the number of zeros that result, in the singularity pattern, from a spontaneous occurence of 0 two steps before, $Z_{n-2}$. Given the form of (\eqxii) the total number of $\infty$ at $n$ is due to a spontaneous occurrence of 0 at the previous step (but notice that the multiplicity of $\infty$ is 2) and those arising from a cyclic pattern that we shall omit. We can thus write the relation
$$Z_n+Z_{n-2}\simeq 2Z_{n-1}.\eqdef\eqxiii$$
The characteristic equation for (\eqxiii) is $(\lambda-1)^2=0$, consistent with the criterion for integrability, namely the absence of a characteristic root greater than 1. Note that (\eqxiii), interpreted as a strict equality, is exactly the same equation as the one satisfied by $a_n$. In fact, it is this observation that lies at the heart of the full-deautonomisation approach [\redemp, \mase].

Where the express method acquires its full usefulness is in the case of non-integrable mappings. We illustrate this with the well-known Hietarinta-Viallet (H-V) [\hiv] mapping:
$$x_{n+1}+x_{n-1}={x_n}+{1\over x_n^2}.\eqdef\eqxiv$$
Its confined singularity pattern is 
$$\{0,\infty^2,\infty^2,0\}.$$
Neglecting possible contributions from cyclic patterns we compute the number of zeros and infinities at each step. Clearly a 0 either occurs spontaneously at step $n$, or as a consequence of a 0 spontaneously occurring three steps before. Similarly an $\infty$ appears, with multiplicity 2, as a consequence of a zero either one or two steps before.  As all other contributions, from cyclic patterns, are bounded, this yields the relation:
$$Z_n+Z_{n-3}\simeq2Z_{n-1}+2Z_{n-2},\eqdef\eqxv$$
the characteristic equation of which factorises to
$$(\lambda+1)(\lambda^2-3\lambda+1)=0.\eqdef\eqxvi$$
The largest root of (\eqxvi) is $(3+\sqrt 5)/2$, which is precisely the dynamical degree of the mapping, obtained in [\take] and [\redemp].

The example above shows that the express method can indeed be used as an integrability detector. So, by performing a simple singularity analysis we can, using this method, set up a linear system the solution of which shows whether exponential growth is present and actually gives the value of the algebraic entropy. 
\bigskip\pagebreak
4. {\scap Some selected applications of the express method}
\medskip
The first mapping we shall examine is a variant of the H-V one, introduced in [\redeem]:
$$x_{n+1}+x_{n-1}=1+{a_n\over x_n^k},\eqdef\eqxvii$$
with $a_{n+4}=-a_n$ for $k\geq 2$, but $a_{n+4}=a_{n+3}+a_{n+1}-a_n$ when $k=1$.
The confined singularity pattern is
$$\{0,\infty^k,1,\infty^k,0\}.$$
The value 0 either occurs spontaneously at step $n$ or is due to a 0 occurring spontaneously four steps before. An $\infty$ appears with multiplicity $k$ as a consequence of a 0 one or three steps before. We can thus write the recursion
$$Z_n+Z_{n-4}\simeq k(Z_{n-1}+Z_{n-3}).\eqdef\eqxviii$$
The characteristic equation is now
$$\lambda^4-k(\lambda^3+\lambda)+1=0,\eqdef\eqxix$$
the largest root of which is $(k+\sqrt{k^2+8})/4+\sqrt{(k\sqrt{k^2+8}+k^2-4)/8}$. This root is greater than 1 for $k>1$ in accordance with the non-integrable character of the mapping, the case $k=1$ corresponding to a well-known integrable case.
Note that in the analysis above we have not considered the occurrences of the value 1. From the singularity pattern one readily sees that a value of 1 may appear when there exists a 0 that appears spontaneously two steps before. However no further information on the possible occurrences of the value 1 can be obtained. Still, one can conclude on the integrability of (\eqxvii) without requiring this information. 

Next we turn to a mapping introduced by Bedford and Kim [\refdef\kim]:
$$x_{n+1}={x_n-a\over x_{n-1}-b}.\eqdef\eqxx$$
When $a\ne0$, $b\ne0,a$, two singularity patterns exist. The first one is
$$\{b,f,\infty,\infty,f',0\},$$
where $f,f'$ are finite values depending on the initial condition $x_0$ and on the parameters of the mapping. A second singularity exists when $x_1=a$. It is generically not confined unless some condition is imposed, namely that at some iteration step, say $m$, we have $x_m=a$ and $x_{m-1}=b$, as shown by Bedford and Kim. The associated pattern, of length $m$, is
$$\{a,0,\cdots,b,a\},$$
where $m\ge4$. Note that a pattern with $m=2$ necessarily entails $a=b=0$, which corresponds to a periodic mapping with period 6. A length-3 pattern can exist provided $b=0$, which is a case that will be studied separately. 

In case $b\ne0$, the value $a$ either occurs spontaneously at step $n$ or is due to a value $a$ occurring spontaneously at step $n-m+1$. Similarly, a value $b$ is either due to a value $a$ occurring spontaneously $m-2$ steps before or it occurs spontaneously. Moreover, as can be seen from the pattern $\{b,f,\infty,\infty,f',0\}$,  a value $\infty$ occurs at step $n$ due to a value $b$ at step $n-2$ or at step $n-3$.

We denote by $A_n$ and $B_n$ the spontaneous occurrences of  $a$ and $b$ at some step $n$ and obtain the relations
$$A_n+A_{n-m+1}\simeq B_n+A_{n-m+2}\simeq B_{n-2}+B_{n-3}
.\eqdef\eqxxi$$ 
Eliminating $B$ we find for $A$ the equation
$$A_n-A_{n-2}-A_{n-3}+A_{n-m+1}+A_{n-m}-A_{n-m-2}\simeq 0,\eqdef\eqxxii$$
the characteristic equation of which is
$$\lambda^3+\lambda^2-1+\lambda^{m-1}(\lambda^3-\lambda-1)=0.\eqdef\eqxxiii$$
Equation (\eqxxiii) is identical to the one obtained for the dynamical degree of (\eqxx) by Bedford and Kim in their algebro-geometric analysis of this mapping [\refdef\bedford].
The first four cases corresponding to $m=4,5,6,7$, as shown by Bedford and Kim [\kim], give periodic mappings. In fact, the periods are exactly given by the solution of the characteristic equation (\eqxxiii). It is interesting to point out here that our method is not {\sl a priori} adapted to the study of periodic mappings. However, if a degree growth had existed our method would have detected it and, thus, its absence signals that the growth of the mapping, is somehow, arrested. 

The two interesting cases correspond to $m=8$ and $m>8$. For $m=8$ the characteristic equation factorises into $(\lambda-1)^3(\lambda+1)(\lambda^2+\lambda+1)(\lambda^4+\lambda^3+\lambda^2+\lambda+1)=0$, which lacks a root greater than 1 and the mapping is therefore integrable. For $m>8$ we remark that, while the value of the characteristic equation at $\lambda=1$ is 0, the corresponding derivative, equal to $8-m$, is always negative. Thus the characteristic equation has at least one root greater than 1, and we are immediately led to the conclusion that the Bedford and Kim mapping is not integrable for $m>8$.

When $m=3$, it is clear that $b$ must be equal to 0, and for the singularity to confine we must also have $a=-1$. However in this case the mapping becomes periodic with period 5. If $b=0$ but $a\neq-1$, confinement does not occur at $m=3$ but at $m=8$. In that case, the fact that $b=0$ makes the first singularity pattern cyclic, and we shall neglect it. The only pattern we are therefore going to consider is 
$$\{a,0,-1,\infty,\infty,-1,0,a\}.$$
We remark that the value $a$ either occurs spontaneously at step $n$ or is due to a value $a$ occurring spontaneously at step $n-7$. Similarly a value 0 at step $n$ is associated either to a value $a$ that occurs at step $n-1$ or at step $n-6$. We obtain thus the equation
$$A_n+A_{n-7}\simeq A_{n-1}+A_{n-6}.\eqdef\eqxxiv$$
Writing the characteristic equation as
$$(\lambda^6-1)(\lambda-1)=0,\eqdef\eqxxv$$
we see that all roots have a modulus equal to 1. Note that instead of 0 we could have considered the contributions to the number of preimages of $\infty$, which would have led to the equation
$$A_n+A_{n-7}\simeq A_{n-3}+A_{n-4},\eqdef\eqxxvi$$
with a different characteristic equation, but again without any roots with modulus greater than 1. The Bedford and Kim mapping for $b=0$ is indeed integrable and its non-autonomous extension has been studied in [\refdef\papy].

Finally we consider the well-known discrete Painlev\'e I equation [\refdef\shohat]
$$x_{n+1}+x_n+x_{n-1}=1+{a_n\over x_n},\eqdef\eqxxvii$$
(where $a_n$ satisfies the relation $a_n-a_{n-1}-a_{n-2}+a_{n-3}=0$) the confined singularity pattern of which is
$$\{0,\infty,\infty,0\}.$$
From the form of the mapping it is clear that any infinite value of $x$ outside the confined pattern could only come from a cyclic pattern. In order to obtain a characteristic equation we equate the contributions, from the spontaneous occurrence $Z_n$ of 0 at step $n$, to the number of preimages of 0 and $\infty$ respectively, leading to
$$Z_n+Z_{n-3}\simeq Z_{n-1}+Z_{n-2},\eqdef\eqxxviii$$
which, incidentally, is exactly the confinement constraint for $a_n$ (when taken as a strict equality).
The characteristic equation for (\eqxxviii) is
$$(\lambda+1)(\lambda-1)^2=0,\eqdef\eqxxix$$
which is, again, in perfect agreement with the integrable character of the mapping. What is interesting in this case is that d-P$_{\rm I}$ possesses longer confined singularity patterns. This is a phenomenon dubbed late confinement in [\mase] and which was originally discussed in [\refdef\HVlate]. The general form of such a pattern is
$$\{0,\infty,\infty,0,\infty,\infty,\cdots,0,\infty,\infty,0\},$$
where the block $(0,\infty,\infty,)$ is repeated $\ell$ times. From the contributions to the number of preimages of 0 and $\infty$ we can set up an equation for $Z_n$. We obtain
$$Z_n-Z_{n+1}-Z_{n+2}+Z_{n+3}+\cdots+Z_{n+3\ell-3}-Z_{n+3\ell-2}-Z_{n+3\ell-1}+Z_{n+3\ell}\simeq0,\eqdef\eqxxx$$
which is exactly the one obtained in [\mase] through an algebrogeometric analysis. It was shown there that for $\ell>1$ a characteristic root greater than 1 always exists and that the late confinement of d-P$_{\rm I}$ leads to non-integrable systems. 
One interesting question one can ask at this level is what happens if one indefinitely postpones confinement, an idea first explored in [\redeem]. Clearly in this case we are talking about a non-confining, non-integrable system. Still, our analysis of the confined integrable system can yield the value of the algebraic entropy of the latter. We start by considering the characteristic equation associated to (\eqxxx) and reorganise it as
$$1=\left({1\over \lambda}+{1\over \lambda^2}-{1\over \lambda^3}\right)\left(1+{1\over \lambda^3}+{1\over \lambda^6}+\cdots+{1\over \lambda^{3\ell-3}}\right).\eqdef\aqxxxi$$
We can sum the series on the right-hand side and, since $\lambda$ is larger than 1, take the limit $\ell\to\infty$. We obtain, after some simplifications, the following equation
$$\lambda^2-\lambda-1=0,\eqdef\aqxxxii$$
the larger root of which is $(1+\sqrt 5)/2$. We have computed numerically the dynamical degree of (\eqxxvii) for generic $a_n$, using the Diophantine Integrability method of Halburd [\refdef\halburd], obtaining after 25 iterations a value of 1.61804, in perfect agreement with the result offered by the infinitely late confinement approach.
\bigskip
5. {\scap The trouble with short singularity patterns}
\medskip
In this section we are going to study an example where a straightforward application of the express method does not lead to useful conclusions and show how, in the present case, this problem can be resolved. Let us consider the mapping
$$\left({x_{n+1}+x_n-z_{n+1}-z_n\over x_{n+1}+x_n}\right)\left({x_{n-1}+x_n-z_{n-1}-z_n\over x_{n-1}+x_n}\right)
={\big((x_n-z_n)^2-a^2\big)\big((x_n-z_n)^2-b^2\big)\over (x_n^2-c^2)(x_n^2-d^2)},\eqdef\eqxxxi$$
which has been studied in detail in [\refdef\oures] where it was shown that for (\eqxxxi) to integrable, $z_n$ must obey the constraint 
$$z_{n+1}-2z_n+z_{n-1}=0,\eqdef\eqxxxii$$ 
(which means that either $z_n$ is constant or it depends linearly on $n$). The 8 singularity patterns of (\eqxxxi)
$$\{\pm c,\mp c\},\ \{\pm d,\mp d\},$$
$$\{z_n\pm a,z_{n+1}\mp a\},\ \{z_n\pm b,z_{n+1}\mp b\},$$
are all of minimal length.
Note that $x_n$ can take at some iteration  the value $\infty$ but when $z_n$ obeys the constraint (\eqxxxii), it turns out that the value $x_n=\infty$ is not a singularity, in the sense that the value of $x_{n+1}$ is perfectly regular. We remark that all 8 singularities play the same role and therefore the number of spontaneous occurrences of each of those singularities  at step $n$ is the same. We denote this quantity by $X_n$. For all these values the total contribution to the number of preimages is $X_n+X_{n-1}$. However, since there is no other quantity we can equate $X_n+X_{n-1}$ to, we become stuck since we do not have a genuine equation for $X_n$. 

Fortunately, a solution to this difficulty exists. We start by introducing an auxiliary variable $y_n$ defined in such a way that equation (\eqxxxi) becomes
$$y_ny_{n-1}={\big((x_n-z_n)^2-a^2\big)\big((x_n-z_n)^2-b^2\big)\over (x_n^2-c^2)(x_n^2-d^2)},\eqdaf\eqxxxiii$$
complemented by
$$x_{n+1}+x_n={z_{n+1}+z_n \over 1-y_n}.\eqno(\eqxxxiii b)$$
The singularity patterns of (\eqxxxiii) are
$$\{x_n=\pm c,y_n=\infty, x_{n+1}=\mp c\},\ \{x_n=\pm d,y_n=\infty, x_{n+1}=\mp d\},$$
$$\{x_n=z_n\pm a,y_n=0, x_{n+1}=z_{n+1}\mp a\},\ \{x_n=z_n\pm b,y_n=0, x_{n+1}=z_{n+1}\mp b\}.$$
Moreover, $x_n$ can take at some iteration  the value $\infty$, provided $y_n$ was equal to 1 at the previous step. This leads to the last singularity pattern
$$\{y_{n-1}=1,x_n=\infty,y_n=1\},$$
which is confined provided the constraint (\eqxxxii) is satisfied. Thanks to the introduction of the auxiliary variable we have now one more singularity pattern and this allows us to obtain a non trivial system of equations. We have, as previously, the quantity $X_n+X_{n-1}$ which gives the sum of the contributions of $X_n$  to the number of preimages, for any of the eight singularities. But now we must also consider the number of preimages of $x_n=\infty$, which is equal to the number of spontaneous occurrences of $y=1$ at the previous step, denoted by $U_{n-1}$. This allows us to set up a first equation
$$X_n+X_{n-1}\simeq U_{n-1}.\eqdef\eqxxxiv$$
Similarly, since a value of $y_n=1$ either appears spontaneously at some step $n$, or is due to a value 1 appearing in the previous step, the number of preimages of $y_n=1$ is  $U_n+U_{n-1}$. On the other hand, a value of $y_n$ equal to 0 or $\infty$ can only appear at some step provided $x_n$ takes any of the four values entering the corresponding singularity pattern. We obtain thus a second equation
$$U_n+U_{n-1}\simeq4X_{n}.\eqdef\eqxxxv$$
Eliminating $X_n$ we obtain for $U_n$ the equation
$$U_{n}-2U_{n-1}+U_{n-2}\simeq0,\eqdef\eqxxxvi$$
leading to a characteristic equation $(\lambda-1)^2=0$. This is in perfect agreement with the integrable character of (\eqxxxi).

At this point it becomes interesting to consider the situation where the constraint (\eqxxxii) is not satisfied. In this case the singularity $y_{n-1}=1,x_n=\infty$ becomes non-confined. However a possibility of confinement at some later step always exists. We may have for instance a first late confinement of the form $\{y_{n-1}=1,x_n=\infty,y_n=1,x_{n+1}=\infty, y_{n+1}=1\}$ provided the constraint $z_{n+3}-2z_{n+2}-2z_{n+1}+z_n=0$ holds. Based on the singularity patterns we then find the system
$$X_n+X_{n-1}\simeq U_{n-1}+U_{n-2},\eqdef\eqxxxvii$$
$$U_n+U_{n-1}+U_{n-2}\simeq4X_{n},\eqdef\eqxxxviii$$
and we can again eliminate $X_n$ obtaining for $U_n$ the equation
$$U_n+U_{n-3}\simeq2(U_{n-1}+U_{n-2}).\eqdef\eqxxxix$$
Note that, once again, this relation coincides with the confinement constraint for the coefficient $z_n$ in this (first) late confinement of the mapping. The characteristic equation for (\eqxxxix) is 
$$\lambda^3-2\lambda^2-2\lambda+1=0,\eqdef\eqxL$$
the largest root of which is $(3+\sqrt 5)/2$ and the late confinement of the mapping is therefore non-integrable.

Just as in the case of the discrete Painlev\'e I (\eqxxvii) we can consider the case where confinement is indefinitely postponed. The characteristic equation for a confinement postponed $\ell-2$ times is
$$1+{1\over\lambda^\ell}=2\left({1\over\lambda}+\cdots+{1\over\lambda^{\ell-1}}\right).\eqdef\eqxLi$$
Given that we know $\lambda$ to be greater than 1 we can sum the series in the right-hand side and take the limit $\ell\to\infty$. We find readily a dynamical degree $\lambda=3$. Computing numerically the dynamical degree, again using Halburd's Diophantine method [\halburd], we find after 10 iterations a value of 3.0003, corroborating the result obtained by the infinitely late confinement.
\bigskip
6. {\scap Conclusions}
\medskip
In this paper we introduced a method which allows us to investigate the integrable character of a given three-point mapping in a particularly simple way, provided the mapping has confined singularities. Our method is inspired by the approach of Halburd who presented a way to compute the degree growth of a given mapping based on the structure of its singularities. The application of Halburd's method necessitates the precise knowledge of all singularity patterns of a given mapping, including those which appear cyclically and which may in some cases be difficult to obtain. Our method, on the other hand, is particularly simple since it uses only the ``standard'' confined singularity patterns of the mapping. The price one pays for this simplification is that one does not have access to the exact value of the degree for the mapping. Still, our method allows  us to compute the dynamical degree and thus to conclude whether a given mapping is integrable or not. Our ``express'' method therefore offers a most straightforward way to compute the algebraic entropy of a mapping.

We presented a collection of examples which illustrated the usefulness of our approach. It turns out that in the case of integrable mappings with a single singularity pattern and a unique parameter to deautonomise, the equation leading to the dynamical degree is the same as the integrability condition on the parameter. It is in fact this observation that was at the heart of the proposal of the full-deautonomisation method, as a discrete integrability criterion. In the case of the non-integrable confining mappings that we  considered, like the H-V mapping and its extensions, our results confirm those obtained in [\redeem] using the full-deautonomisation method. On the other hand the present method allows us to deal with mappings, like the one of Bedford and Kim, where deautonomisation calculations become prohibitively bulky.

While particularly powerful, the express method may face a problem when all singularity patterns are very short, i.e. when the singularity is confined in one step, in which case one cannot write an equation leading to the dynamical degree. Still, as we have shown in the example of section 5, it is possible to resolve this difficulty by introducing an auxiliary variable. It remains to be seen, however, if such a solution is possible in every case where the singularity patterns are too short. 

An interesting result, already obtained by the full-deautonomisation approach [\redeem] but equally and sometimes more easily  accessible with the express method, is that one can use results from a confining system in order to obtain information on a non-confining one. The trick is, starting from a mapping with confined singularities, to deautonomise it while postponing confinement. An indefinitely delayed confinement corresponds to a non-integrable system, with unconfined singularities, for which we can still compute the dynamical degree as a limit. In the examples of infinitely late confinement presented in this paper we have also confirmed our results by a direct computation of the dynamical degree.

Thus, we can conclude that the express method is a powerful discrete integrability detector, giving a simple ``yes or no'' answer to the question of the integrability of a mapping. When the system is integrable one can then proceed to its deautonomisation, in order to obtain its full freedom, deploying the singularity arsenal.
\bigskip
{\scap Acknowledgements}
\medskip
The authors are greatly indebted to Rod Halburd who graciously provided them with an early version of his paper and who allowed them to forge ahead with their own manuscript.
RW would like to acknowledge support from the Japan Society for the Promotion of Science (JSPS), through the the JSPS grant: KAKENHI grant number 15K04893. TM would also like to acknowledge support from JSPS through the grant: KAKENHI grant number 16H06711.
\bigskip
{\scap References}
\medskip
\begin{description}
\item{[\veselov]} A.P. Veselov, Comm. Math. Phys. 145 (1992) 181.
\item{[\arnold]} V. I. Arnold, Bol. Soc. Bras. Mat. 21 (1990) 1.
\item{[\bellon]} M. Bellon and C-M. Viallet, Comm. Math. Phys. 204 (1999) 425.
\item{[\scae]} Y. Ohta, K.M. Tamizhmani, B. Grammaticos and A. Ramani, Phys. Lett. A 262 (1999) 152.	
\item{[\sincon]} B. Grammaticos, A. Ramani and V. Papageorgiou, Phys. Rev. Lett. 67 (1991) 1825.
\item{[\hiv]} J. Hietarinta and C-M. Viallet, Phys. Rev. Lett. 81, (1998) 325.
\item{[\redemp]} A. Ramani, B. Grammaticos, R. Willox, T. Mase and M. Kanki, J. Phys. A 48 (2015) 11FT02.
\item{[\mase]} T. Mase, R. Willox, B. Grammaticos and   A. Ramani, Proc. Roy. Soc. A 471 (2015) 20140956.
\item{[\redeem]} B. Grammaticos, A. Ramani, R. Willox, T. Mase and J. Satsuma, Physica D 313 (2015) 11.
\item{[\favre]} J. Diller and C. Favre, Amer. J. Math. 123 (2001) 1135.
\item{[\take]} T. Takenawa, J. Phys. A 34 (2001) 10533.
\item{[\rod]}   R.G. Halburd, ``{\sl  Elementary exact calculations of degree growth and entropy for discrete equations}'', preprint (2016).
\item{[\fokas]} A. Fokas, B. Grammaticos and A. Ramani, J. of Math. Anal. and Appl. 180 (1993) 342.
\item{[\kim]} E. Bedford and K. Kim, Michigan Math. J. 54 (2006) 647.
\item{[\bedford]} E. Bedford and K. Kim, J. Geom. Anal. 19 (2009) 553.
\item{[\papy]} M.D. Kruskal, K.M. Tamizhmani, B. Grammaticos and A. Ramani, Regul. Chaotic Dyn. 5 (2000) 273.
\item{[\shohat]} J.A. Shohat, Duke Math. J. 5 (1939) 401.
\item{[\HVlate]} J. Hietarinta and C. Viallet, Chaos, Solitons Fractals 11 (2000) 292.
\item{[\halburd]} R.G. Halburd, J. Phys. A 38 (2005) L263.
\item{[\oures]} A. Ramani, R. Willox, B. Grammaticos, A.S. Carstea and J. Satsuma, Physica A 347 (2005) 1.
\end{description}

\end{document}